\documentclass[preprint,12pt]{elsarticle}

\usepackage{amssymb}

 \def\gsim{\mathrel{\rlap{\lower4pt\hbox{\hskip1pt$\sim$}}
 \raise1pt\hbox{$>$}}}

 \newcommand\la{\langle}
 \newcommand\ra{\rangle}
 \newcommand\beq{\begin{equation}}
 
 \newcommand\eeq{\end{equation}}
 \newcommand\beqn{\begin{eqnarray}}
 \newcommand\eeqn{\end{eqnarray}}

\def\fm{\,\mbox{fm}}
\def\GeV{\,\mbox{GeV}}
\def\TeV{\,\mbox{TeV}}
\def\lsim{\mathrel{\rlap{\lower4pt\hbox{\hskip1pt$\sim$}}
    \raise1pt\hbox{$<$}}}         
\def\gsim{\mathrel{\rlap{\lower4pt\hbox{\hskip1pt$\sim$}}
    \raise1pt\hbox{$>$}}}         

\def\fm{\,\mbox{fm}}
\def\GeV{\,\mbox{GeV}}

\def\beq{\begin{equation}}
\def\eeq{\end{equation}}

\def\beqy{\begin{eqnarray}}
\def\eeqy{\end{eqnarray}}

\journal{Physics Letters B}

\begin{document}

\begin{frontmatter}

\title{Mutual boosting of the saturation scales\\ in colliding nuclei}

\author{B. Z. Kopeliovich$^{1,2}$, H. J. Pirner$^{2}$, I. K. Potashnikova$^{1}$, and Iv\'an Schmidt$^{1}$}

\address{$^{1}$Departamento de F\'{\i}sica, 
Universidad T\'ecnica Federico Santa Mar\'{\i}a; and
\\
Instituto de Estudios Avanzados en Ciencias e Ingenier\'{\i}a; and\\
Centro Cient\'ifico-Tecnol\'ogico de Valpara\'iso;\\
Casilla 110-V, Valpara\'iso, Chile}
\address{$^2$Institut f\"ur Theoretische Physik der Universit\"at,\\
Philosophenweg 19, 69120 Heidelberg, Germany}
\begin{abstract}

Saturation of small-$x$ gluons in a nucleus, which has the
form of transverse momentum broadening of projectile gluons in $pA$
collisions in the nuclear rest frame, leads to a modification of the
parton distribution functions in the beam compared with $pp$
collisions. The DGLAP driven gluon distribution turns out to be
suppressed at large $x$, but significantly enhanced at $x\ll1$. This
is a high twist effect. In the case of nucleus-nucleus collisions
all participating nucleons on both sides get enriched in gluon
density at small $x$, which leads to a further boosting of the
saturation scale. We derive reciprocity equations for the saturation
scales corresponding to a collision of two nuclei. The solution of
these equations for central collisions of two heavy nuclei
demonstrate a significant, up to several times,  enhancement of
$Q_{sA}^2$, in $AA$ compared with $pA$ collisions.
\end{abstract}


\end{frontmatter}

\section{Introduction}

The transverse momentum distribution of gluons in nuclei is known to
be modified compared with a free nucleon. The mean transverse
momentum squared increases up to a value called saturated scale,
$Q_{sA}^2$, which depends on the nuclear profile. This phenomenon,
called color glass condensate \cite{mv}, is related to parton
saturation at small $x$ \cite{glr}, and can be also understood in
terms of the Landau-Pomeranchuk principle \cite{lp} as a consequence
of coherent gluon radiation from multiple interactions in the
nucleus \cite{broad}. The value of the saturation momentum was
calculated and compared with data on broadening in \cite{broad} and
has been modeled recently in
\cite{raju1,kh1,kh2,kh3,kov1,kov2,w2,w3}.

The saturation scale can be measured as $p_T$-broadening of a parton
propagating through the nucleus  in its rest frame \cite{broad},
\beq
Q_{sA}^2=\Delta p_T^2. 
\label{10}
\eeq
Although in leading order both sides of this relation rise linearly
with nuclear profile $T_A$ \cite{pir,dhk,jkt,bdmps,raju}, this
dependence slows down by gluon shadowing. 
These phenomena,
broadening and suppression of gluons, are closely related, since
both result from coherence of gluon radiation in multiple
interactions. Solving the corresponding equation derived in
\cite{broad}, one arrives at a saturation scale considerably reduced
compared to the leading order. The $T_A$ dependence is slower than
linear, and at very large (unrealistic) nuclear thicknesses the
saturation scale saturates, becoming independent of $T_A$. Notice
that the solution found in \cite{broad} is similar to the result of
numerical solution  \cite{w1} of the Balitsky-Kovchegov equation \cite{b,k}.
Broadening of gluons radiated in heavy ion collisions was studied with numerical simulations in \cite{kv,lappi}.

\section{Modification of the beam PDF by a nuclear target}

Due to broadening a nuclear target probes the parton distribution in the beam hadron with a higher resolution. Therefore, the effective scale $Q^2$ for the beam PDF drifts to a higher value $Q^2+Q_{sA}^2$. At first glance this seems to contradict casuality, indeed, how can the primordial parton distribution in the hadron depend on the interaction which happens later? However, there is nothing wrong. The interaction performs a special selection of Fock states in the incoming hadron. The same phenomenon happens when one is measuring the proton parton distribution in DIS. The proton PDF "knows" in advance about the virtuality of the photon which it is going to interact with.

The shift in the scale also can be interpreted as a manifestation of the Landau-Pomeranchuk principle \cite{lp}: at long coherence times gluon radiation (which causes the DGLAP evolution) does not depend on the details of multiple interactions, but correlates only with the total momentum transfer, $\vec q+\Delta\vec p_T$, which after squaring and averaging over angles results in $Q^2+\Delta p_T^2$.

As far as the PDF of the projectile proton has a harder scale in $pA$ collisions than in $pp$, the ratio of parton distributions should fall below one at forward and rise above one at backward rapidities. This may look like a breakdown of $k_T$-factorization, however, it is a higher twist effect. 

\begin{figure}[htb]
 \includegraphics[width=6cm]{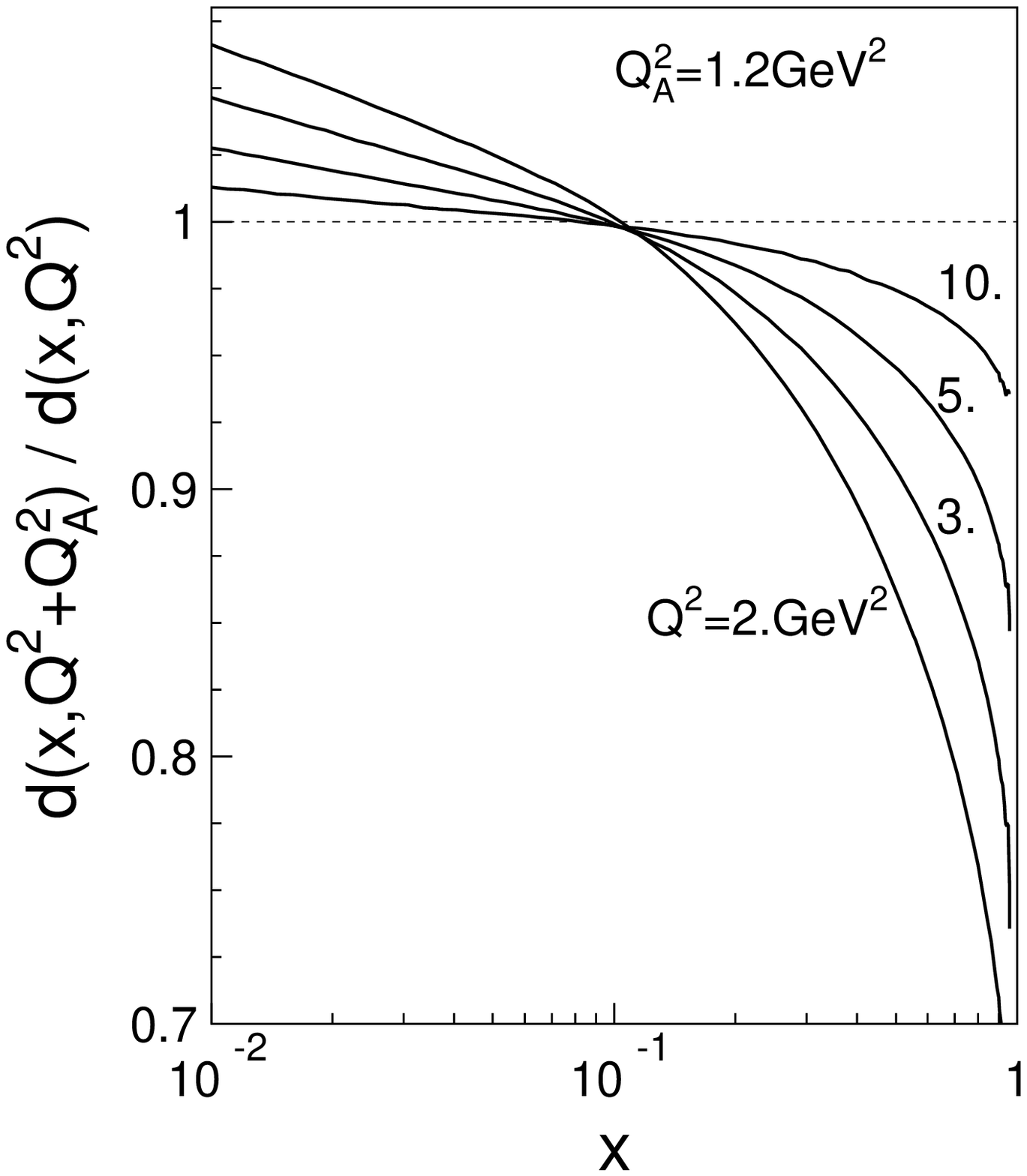} \hspace{5mm}
 \includegraphics[width=6cm]{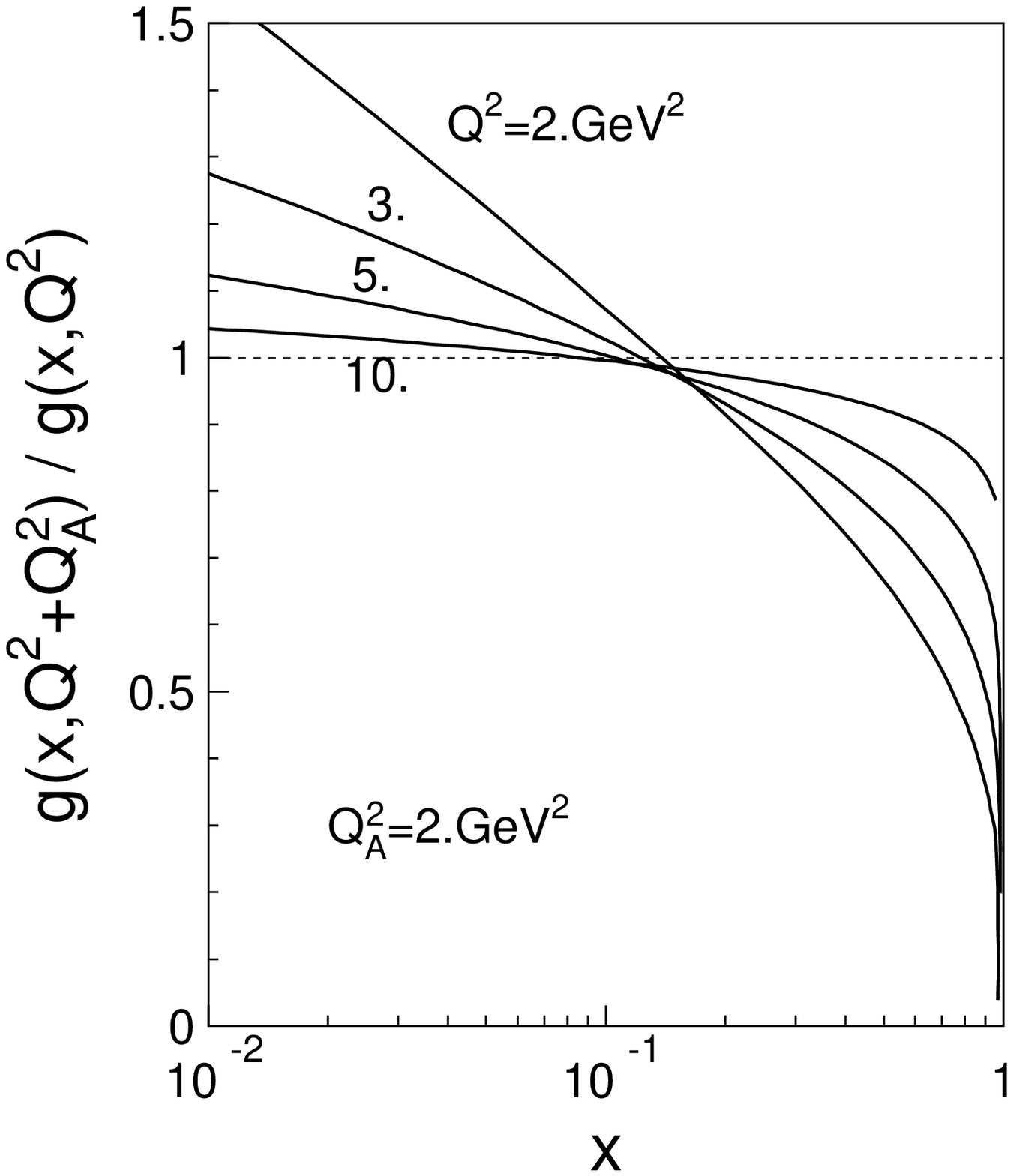}
 \caption{\label{rescaling} Ratio of parton distribution functions in a reaction characterized by a hard scale $Q^2=2,\ 3,\ 5,\ 10\GeV^2$ on a nuclear (A=200) and proton targets.   {\it Left panel:} ratio of the $d$-quark distributions for the quark saturation momentum $Q_{sA}^2=1.2\GeV^2$.
{\it Right panel:} Ratio for gluons with the gluon saturation momentum $Q_{sA}^2=2\GeV^2$.}
 \end{figure}

Examples of $pA$ to $pp$ ratios $R_A(x,Q^2)$ calculated with MSTW2008 \cite{mstw} are shown in Fig.~\ref{rescaling} for $d$-quark and gluon distributions in a hard
reaction (high-$p_T$, heavy flavor production, etc.). 
We see that the shift in the hard scale caused by saturation in the
nucleus leads to a sizable suppression in the projectile parton
distribution at large $x\to1$ and enhancement at small $x\ll1$. We
also observe that the magnitude of nuclear modification quickly decreases with $Q^2$ confirming that this is a high twist effect.

Important for what follows is the observation of a considerably increased  population
of small-$x$ partons in the projectile proton in $pA$ compared with $pp$ collisions.

\section{Nucleus-nucleus collisions: reciprocity relations}
Notice that in $pA$ collisions the modification of the PDFs of the beam and target are not symmetric. Namely, the scale of the PDF of the beam proton gets a shift, $Q^2\Rightarrow Q_{eff}^2=Q^2+Q_{sA}^2$,
while the PDFs of the bound nucleons, which do not undergo multiple interactions, remain the same as in $pp$ collisions.

\begin{figure}[htb]
\begin{center}
 \includegraphics[width=10cm]{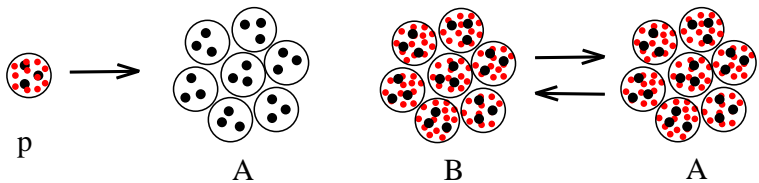}
 \end{center}
\caption{\label{pA-AA} {\it Left:} $pA$ collision in which the
colliding proton is excited by multiple interactions up to a
saturated scale $Q_{sA}^2$, what leads to an increased multiplicity
of soft gluons in the incoming proton. {\it Right:} nuclear
collision in which participating nucleons on both sides are boosted
to the saturation scales, $Q_{sA}^2$ in the nucleus $B$, and
$Q_{sB}^2$ in the nucleus $A$. As a result, the low-$x$ gluon
population is enriched in both nuclei.}
 \end{figure}

The situation changes in the case of a nucleus-nucleus collision: the bound
nucleons in both nuclei participate in multiple interactions,
therefore the scales of PDFs of all of them are modified. However,
this modification goes beyond the simple shift $Q^2\Rightarrow
Q^2+Q_{sA}^2$. Indeed, in an $AB$ nuclear collision not only the two
nucleons (one from $A$ and one from $B$) participating in the hard
reaction undergo multiple interactions, but also many other
nucleons, the so called participants, experience multiple soft
interactions. For this reason their parton distributions are boosted
from the soft scale $\mu^2$ up to the saturation scale
$\mu^2\Rightarrow \mu^2+Q_{sA(B)}^2$, which is usually much larger.
Thus, the participant nucleons on both sides are boosted to a higher
scale and get softer PDFs, with larger parton multiplicities at
small $x$. This is illustrated on the cartoon in Fig.~\ref{pA-AA}.

The next important observation is that the $p_T$-broadening on such
"excited", or boosted nucleons, $\tilde N$ is larger than in $pA$ collisions,
$\Delta p_T^2\bigr|_{\tilde N}>\Delta p_T^2\bigr|_N$, since the density 
of target gluons is increased at small $x$. This should lead to
a further mutual enhancement of broadening, i.e. a further increase of
the saturation scales in both nuclei. Intuitively this seems to be clear, but
a formal consideration below also supports this conclusion.

Broadening is predominantly a process based on many soft
rescatterings of the projectile parton. It was found in
\cite{jkt,dhk} (see also \cite{broad}) that quark broadening is related to the
dipole cross section, 
\beq 
Q_{sA}^2=\Delta p_T^2(E) =
2\frac{d\sigma(r,E)}{dr^2}\Bigr|_{r=0}\,\int dz\,\rho_A(b,z),
\label{410} 
\eeq 
where $\rho_A(b,z)$ is the nuclear density at
impact parameter $b$ and longitudinal coordinate $z$, Since the
process is soft, Bjorken $x$ is not a proper variable, but instead
the parton energy $E$ should be used. The energy dependent $\bar qq$ dipole
cross section was parametrized in the saturated form and fitted to photoabsorption, low
$Q^2$ DIS data, and $\pi p$ total cross section in \cite{kst2} (see also \cite{jkt,hikt1}). With that
parametrization  \cite{jkt} 
\beq 
C_q(E) \equiv
\frac{d\sigma(r,E)}{dr^2}\Bigr|_{r=0}={1\over4}\, \sigma^{\pi
p}_{tot}(E)\left[Q_{qN}^2(E)+\frac{3}{2 \la r_{ch}^2\ra_\pi}\right],
\label{420} 
\eeq 
where $\sigma^{\pi p}_{tot}(E)$ is the $\pi p$
total cross section; $\la r_{ch}^2\ra_\pi\approx 0.44\fm^2$ is the
mean pion charge radius squared; $Q_{qN}(E) =
0.19\GeV\times(E/1GeV)^{0.14}$ is the proton saturation momentum.

Notice, that Eq.~(\ref{420}) $C_q(E)$  has no scale dependence.
It corresponds to the dipole-nucleon cross section
$\sigma_{dip}(r)=C_q\,r^2$ and a soft scale characterizing the proton is implicitly contained in $C_q$. Strictly speaking, however,  this
coefficient is divergent at $r\to0$, since it contains $\ln(r_T)$
\cite{zkl}. This divergency originates from the ultraviolet behavior
of the unintegrated gluon density ${\cal F}(x,k_T)\propto 1/k_T^4$
at large $k_T$. In reality this divergency is not harmful due to the
natural cut-offs discussed in \cite{jkt}, and to a low sensitivity
to their values. One should fix the $\ln r$ dependent factor term in $C_q$ at some
value of $r$ typical for the process under consideration, as is done
in Eq.~(\ref{420}).

The function $C_q(E)$ in (\ref{420}) describes broadening in $pA$ collisions, resulting from multiple interactions with "normal" target nucleons, whose parton distributions are the same as in $pp$ interactions. However,  as was discussed above,
the wave functions of participant nucleons in $AA$ collisions are biased towards a larger scale $Q_s^2$ and a higher multiplicity of the constituent partons. This is why the $C_q$ in (\ref{420}) acquires the second variable, a scale $Q^2=Q_s^2$,  $C_q(E)\Rightarrow C_q(E,Q^2)$. To expose this scale dependence explicitly, we present the function $C_q(E,Q^2)$ in the form \cite{frs,bartels},
\beq
C_q(E,Q^2)=\frac{\pi^2}{3}\,\alpha_s(Q^2)\,xg_N(x,Q^2),
\label{360}
\eeq
where $g_N(x,Q^2)$ is the gluon distribution function in the target
nucleon, and $x=Q^2/2m_NE$. 

One should be careful bridging Eq.~(\ref{360}) with its soft limit Eq.~(\ref{420}),
since  $\alpha_s(Q^2)$ and
$g_N(x,Q^2)$ are ill defined at small $Q^2$. To regularize this
problem we replace $Q^2\Rightarrow Q^2+Q^2_0$, where $Q_0^2$ should
be adjusted to the reproduction of the correct infra-red limit,
\beq 
\frac{\pi^2}{3}\,\alpha_s(Q_0^2)\,xg_N(x,Q_0^2)= C_q(E=Q_0^2/2m_Nx).
\label{370}
\eeq
Apparently, the value of $Q_0^2$ is not universal.
It depends on $s$, $x$, and most of all on the PDF-analysis
dependent behavior of  $g_N(x,Q_0^2)$, especially at small $Q^2$. Here we
follow this procedure.

Thus, a participating nucleon simultaneously plays the roles of a
beam and of a target. As a beam hadron its PDF is boosted to a
higher scale due to multiple interactions it undergoes in another
nuclei. As a target such a nucleon, being boosted to a higher scale
$Q_0^2\Rightarrow Q_0^2+Q_{sB}^2$, it increases broadening of
partons from another nucleus, since the factor  $C_q(E,Q^2)$
Eq.(\ref{360}) rises. This leads to a mutual enhancement of the
saturation scales in both nuclei. Indeed,  multiple rescatterings of
nucleons from the nucleus $A$ on the  boosted nucleons in $B$
proceed with a larger cross section, so broadening, i.e. the
saturation scale in $B$ increases, $Q_{sB}^2\Rightarrow \tilde
Q_{sB}^2>Q_{sB}^2$. For this reason, the nucleon PDFs in $A$ get
boosted more. Then the partons from $B$ experience even stronger
multiple interactions with such double-boosted nucleons in $A$. This
results in an additional boost of the saturation scale in $A$, then, as a result, in $B$, and so on. 

\begin{figure}[htb]
\begin{center}
 \includegraphics[width=12cm]{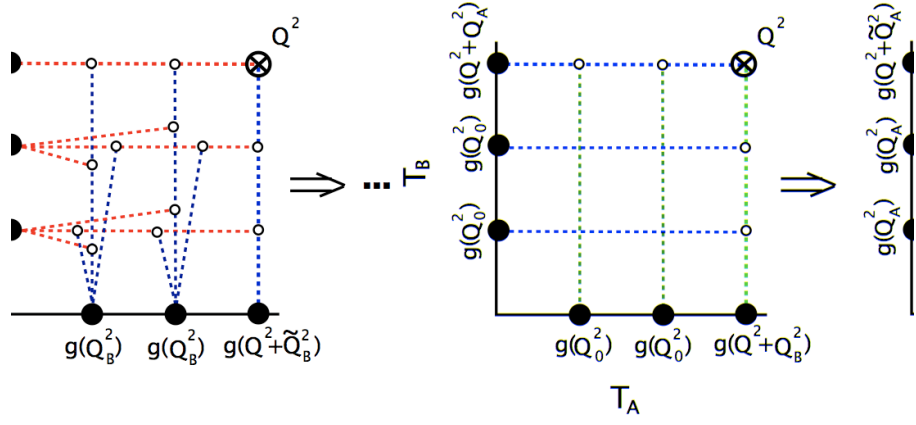}
 \end{center}
\caption{\label{boosting} Collision of two one-dimensional rows of nucleons $T_A$ and $T_B$ displayed on horizontal and vertical axes. 
Multiple interactions of colliding gluons propagated through both nuclei,
including additional multiple scatterings of gluons, which carry out the interactions, are shown as is described in text. }
 \end{figure}

Such a multi-iteration mutual boosting of the saturation scales is illustrated
pictorially in Fig.~\ref{boosting}, where two raws of nucleons, $T_A$ and $T_B$ are displayed on horizontal and vertical axes. 
In the left picture a gluon originated from one of the nucleons in $A$ (the rightmost vertical dashed line) propagates through $B$ interacting multiply via gluon exchanges (horizontal dashed lines) and increasing its transverse momentum squared by $\Delta p_T^2=Q_{sB}^2$. In a similar way a gluon from $B$ (the upper red line) interacts multiply propagating through $A$ and gets broadening $Q_{sA}^2$.  All multiple interactions occur at the soft scale $Q_0^2$.
The two gluons collide (the cross in the upper right corner) at hard scale $Q^2$,
but the original gluon distributions in the colliding nucleons are shifted to higher scales, $Q^2+Q_{sA}^2$ and $Q^2+Q_{sB}^2$ respectively.

Then we observe that the $t$-channel gluon exchanges (vertical lines) between the gluon  originated from $B$ and nucleons bound in $A$, become $s$-channel gluons propagating through $B$, after a Lorentz boost between the rest frames of nuclei $A$ and $B$. Therefore, such gluons should also undergo multiple collisions with nucleons in $B$. This is shown by additional horisonal gluon lines in the right picture of Fig.~\ref{boosting}. And vice versa, the original gluonic exchanges carrying out multiple interactions in the rest frame of $B$, propagate through $A$ and also experience new multiple interactions, as is depicted in the right hand side of Fig.~\ref{boosting}. These additional interactions boost each of the multiple interactions to a new scale, as well as the scale of the hard reaction $Q_{sA(B)}^2\Rightarrow\tilde Q_{sA(B)}^2$.

In the next iteration (not shown in Fig.~\ref{boosting}) the new exchanged gluons also experience multiple scatterings, and so on. 
The final gluon saturation scales $\tilde Q_s^2$ in the collision of two rows of nucleons  $T_A$ and $T_B$ can be found solving the reciprocity equations ($C_q\Rightarrow C_g={9\over4}C_q$),
 \beqn
\tilde Q_{sB}^2(x_B)&=&\frac{3\pi^2}{2}\,\alpha_s(\tilde Q_{sA}^2+Q_0^2)\,
x_B g_N(x_B,\tilde Q_{sA}^2+Q_0^2)\,T_B;
\nonumber\\
\tilde Q_{sA}^2(x_A)&=&\frac{3\pi^2}{2}\,\alpha_s(\tilde Q_{sB}^2+Q_0^2)\,
x_Ag_N(x_A,\tilde Q_{sB}^2+Q_0^2)\,T_A.
\label{380}
\eeqn
These equations are the main result of the paper. Compared to
Eq.~(\ref{420}) these equations take into account the modification
of the properties of bound nucleons in each of the colliding nuclei
due to multiple interactions in another nucleus and the following
increase of the scale. Notice that the values of $Q_0$ in the first
and the second equations (\ref{380}) depend on $x_A$ and $x_B$
respectively, and therefore might be slightly different, but we use
the same symbols to simplify the notations.

The reciprocity equations should be solved numerically, but here we
estimate the magnitude of the effect for the case of  central collision of identical nuclei, i.e. $T_A=T_B$,
and a glue-glue collision at mid-rapidity, $x_A=x_B=x$. In this case the system
of equations (\ref{380}) reduces to a single one,
\beq
\tilde Q_{sA}^2(x)=\frac{3\pi^2}{2}\,\alpha_s(\tilde
Q_{sA}^2+Q_0^2)\, x g_N(x,\tilde Q_{sA}^2+Q_0^2)\,T_A.
\label{400}
\eeq
The  scale characterizing multiple interactions of gluons is the mean transverse momentum of gluons $\la k_T\ra\approx 0.65\GeV$ \cite{kst2,spots}.
Therefore, the gluon distribution should be taken at  $x=\la k_T\ra/\sqrt{s}$, which
gives $x=3.25\times 10^{-3}$ and $x=1.18\times 10^{-4}$, corresponding
to $\sqrt{s}=200\GeV$ (RHIC) and $5.5\TeV$ (LHC) respectively.

To proceed further we should fix the infra-red cutoff $Q_0^2$ given by Eq.~(\ref{370}). At the energy of RHIC the parameter
in the right-hand side of Eq.~(\ref{370}) $C(E=65\GeV)=3.2$. Then  Eq.~(\ref{370}) results in  $Q_0^2(\sqrt{s}=200\GeV)=1.84\GeV^2$.
At the energy of LHC  the factor $C(E=1.787\TeV)=7.03$.
So we found $Q_0^2(\sqrt{s}=5.5\TeV)=1.7\GeV^2$.
These figures confirm our expectation of a weak energy dependence of the infra-red cutoff $Q_0$.

\begin{figure}[htb]
 \includegraphics[width=6cm]{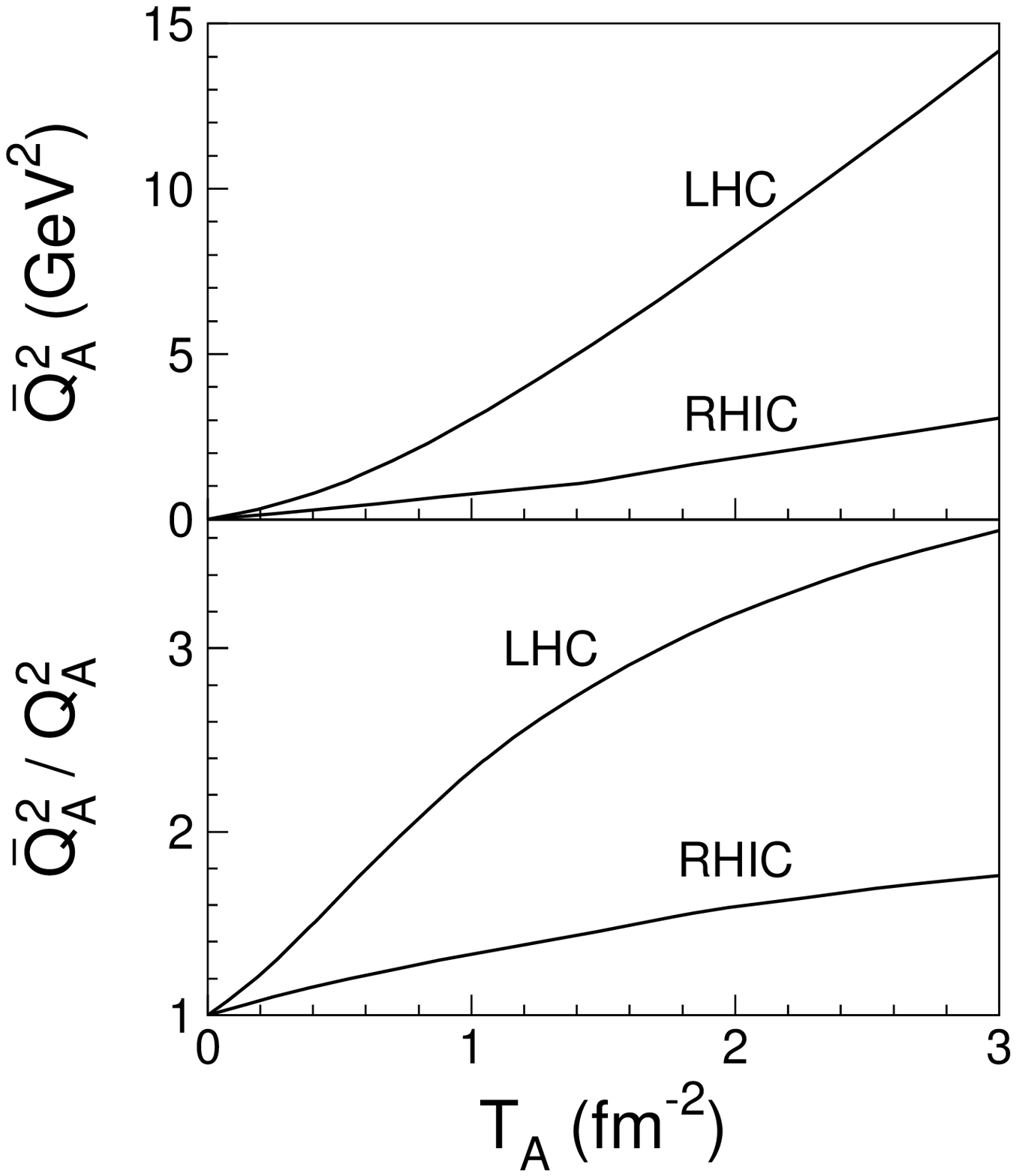}
 \hspace{10mm}
 \includegraphics[width=5cm]{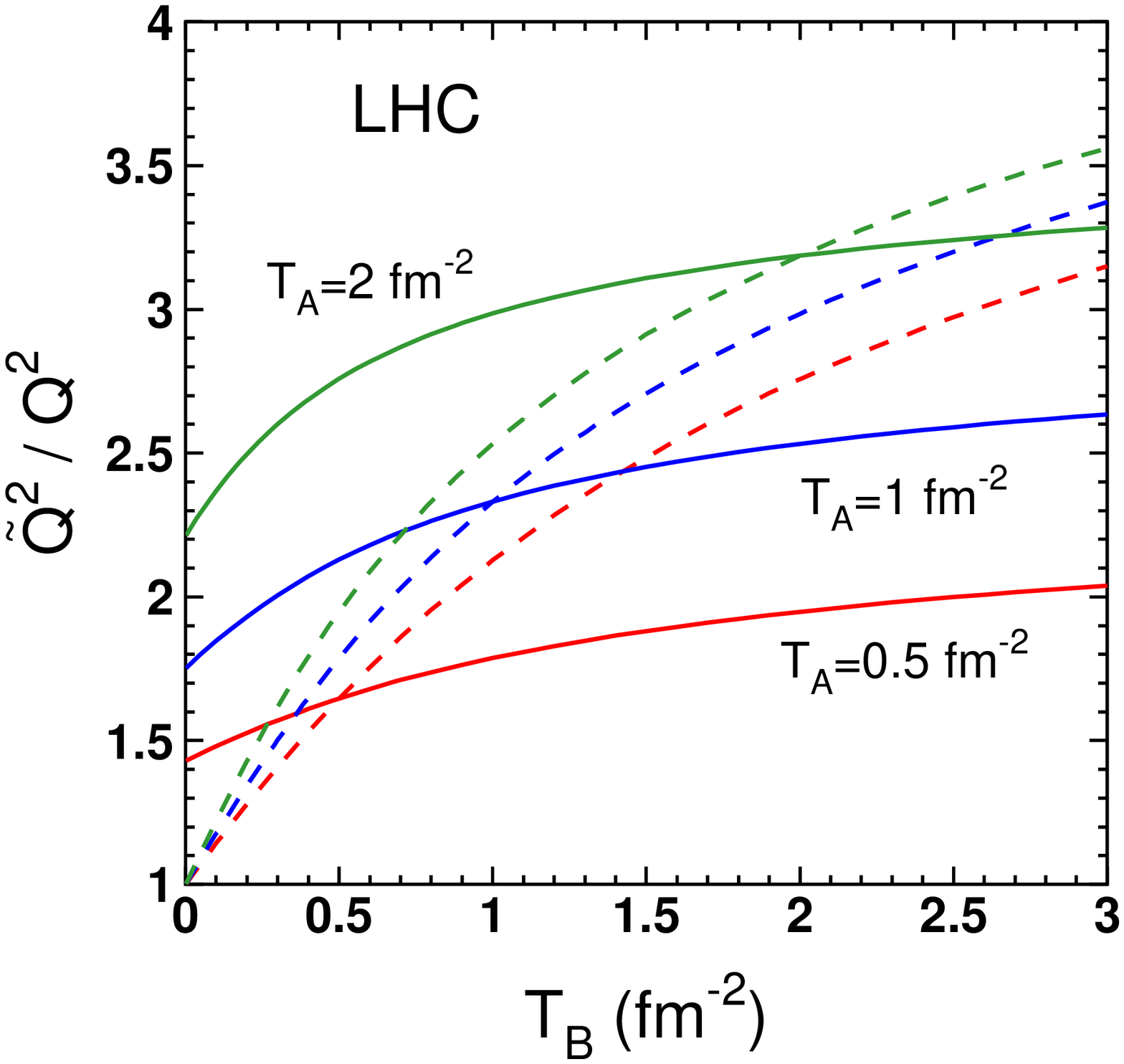}
 \caption{\label{results} {\sl Left upper panel:}  the boosted values of the saturation momentum squared $\tilde Q_{sA}^2$ calculated for $T_A=T_B$ with Eq.~(\ref{400}) at the energies of RHIC and LHC as function of nuclear thickness.
 {\sl Left bottom panel:} the boosting factors $\tilde Q_{sA}^2/Q_{sA}^2$ as function of nuclear thickness.
 {\sl Right panel:} The boosting factors as function of $T_B$ for fixed values of $T_A=2,\,1,\ 0.5\fm^{-2}$ from bottom to upper curves respectively. Solid and dashed curves correspond to the boosting factors for nuclei $B$ and $A$ respectively. }
 \end{figure}

With these values of $x$ and $Q_0$ we solved the equation (\ref{400}) for a central collision of identical nuclei, relative to the modified value of the saturation momentum $\tilde Q_{sA}$  as function of $T_A=T_B$ using the LO gluon distributions of the recent analysis MSTW2008 \cite{mstw}. The results are plotted in the left upper panel of Fig.~\ref{results} as function of nuclear thickness $T_A$ at the energies of RHIC and LHC.
We see that the saturation scale of heavy nuclei may be as large as
about $10\GeV^2$ at the LHC.

To see the magnitude of the boosting effect we also show in the left bottom panel
the boosting factor as function of $T_A=T_B$ at the energies of RHIC and LHC. The enhancement is significant, especially at the
energy of LHC, where it reaches a factor of three.

We also solved the reciprocity equations (\ref{380}) for non-central collisions, i.e. for $T_A\neq T_B$.
The boosting factor is plotted in the right panel of Fig.~\ref{results} as function of $T_B$ for fixed values of $T_A=2,\,1,\ 0.5\fm^{-2}$ from bottom to upper curves respectively. Solid and dashed curves show the boosting factor for nuclei $B$ and $A$ respectively. These results confirm the expectation illustrated pictorially in the left panel of Fig.~\ref{pA-AA}. Namely, in the limit $T_B\to0$, corresponding to $pA$ collisions, the boosting effect in nucleus $A$ vanishes, 
while the parton distribution in the projectile proton is drifting to a higher scale, i.e. the boosting factor exceeds unity. 
At larger values of $T_B$ the numerical results in Fig.~\ref{results} show that the saturation scales in both nuclei 
are boosted to higher values in accordance with Fig.~\ref{pA-AA}, right.

\section{Gluon shadowing}

The gluon density at small $x$ in nuclei is expected to be reduced
compared with free nucleons. This phenomenon called gluon
shadowing, is a part of saturation, but it also affects and
diminishes the saturation scale. With this observation we formulated an
equation for the modified saturation scale in $pA$ collisions
\cite{broad}. The found reduction of the saturation momentum is
significant and is similar to the result of numerical solutions
\cite{w1} of the Balitsky-Kovchegov equation \cite{b,k}. Thus, both
$Q_{sA}^2$ and $\tilde Q_{sA}^2$ plotted in the upper panel of
Fig.~~\ref{results} would be reduced in accordance with
Ref.~\cite{broad} (see Fig.~3 of that paper).

However, the modification of the saturation scale due to mutual boosting in heavy ion collisions turns out to be practically unaffected by gluon shadowing.
Indeed, either the usual saturation scale Eq.~(\ref{360}), or
the reciprocity equations (\ref{380}) are controlled by the gluon distribution function, which should be modified in a nucleus by a factor $R_g(x,Q^2,T_A)$, which is the nucleus-to-nucleon ratio of the gluon PDFs. The difference between the two is in the scales
for gluons shadowing, which are $Q_{sA}^2$ and $\tilde Q_{sA}^2$ respectively. However, the scale dependence of $\alpha_s(Q^2)\,g(x,Q^2)$ at small $x$, given by the DGLAP evolution,  is very slow and can be neglected. Thus, the boosting factors depicted in the bottom panel of Fig.~\ref{results} are not affected by gluon shadowing.

\section{Experimental observables}

Increase of the saturation scales in colliding nuclei should certainly lead to a rise of transverse momenta of produced hadrons in comparison with an extrapolation from $pA$ collisions. However, final state interactions with the dense medium created in heavy ion collisions, significantly modify the transverse momentum distribution of produced hadrons.
Exclusion is production of heavy quarkonia. Propagating through a dense medium they experience no energy loss,
and their survival probability is practically independent of $p_T$. Therefore any observed modifications of the $p_T$-distribution of heavy quarkonia produced in nuclear collisions should be associated with initial state interactions.
This fact makes them an excellent for study of gluon broadening in colliding nuclei.

We expect an increased magnitude of broadening $J/\Psi$ and $\Upsilon$ produced in $AA$ compared with $pA$ collisions for the same path length in nuclear medium. This could be interpreted as a signal of the effect of boosted saturation scale discussed here.

\begin{figure}[htb]
\begin{center}
\includegraphics[width=4.3cm]{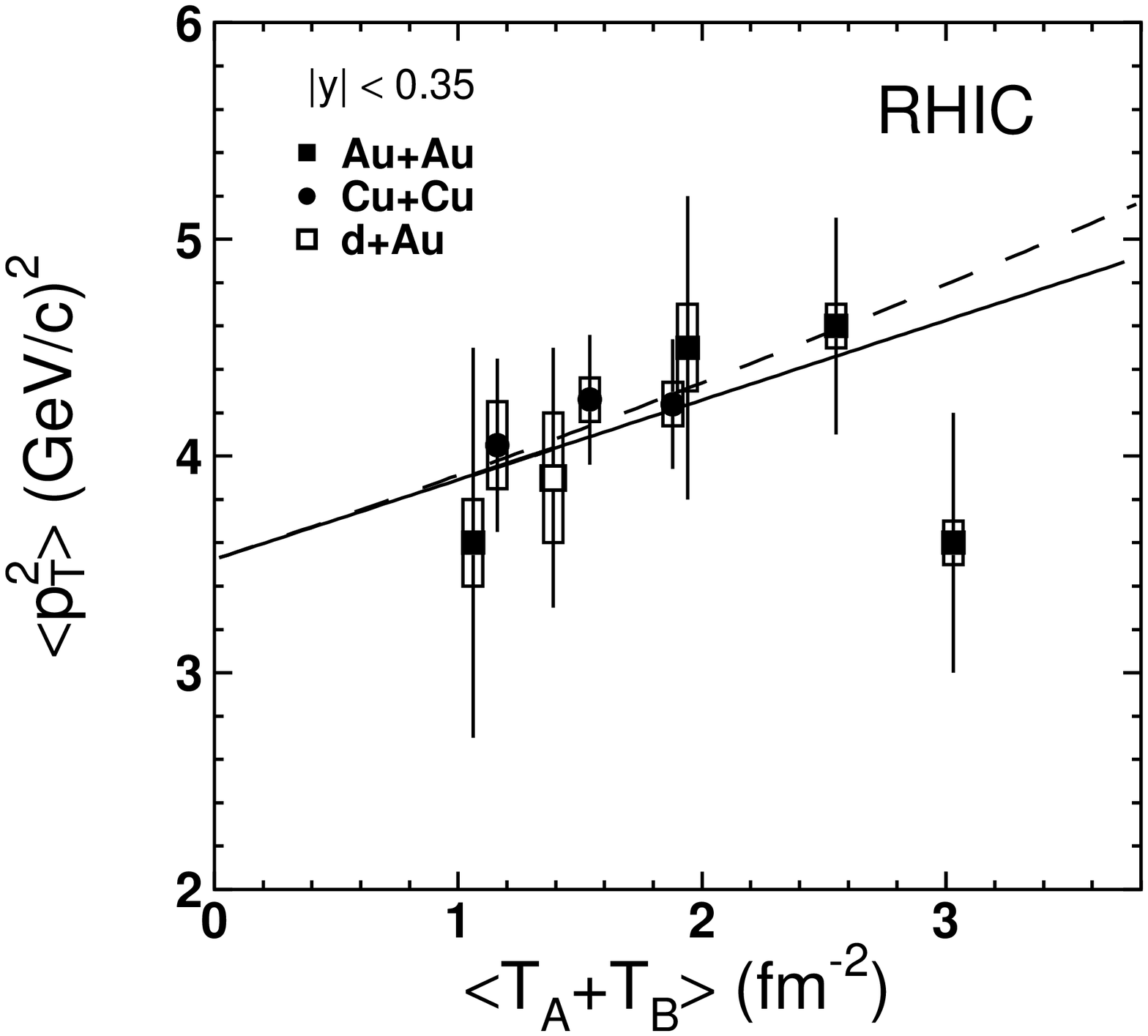}
\includegraphics[width=4.3cm]{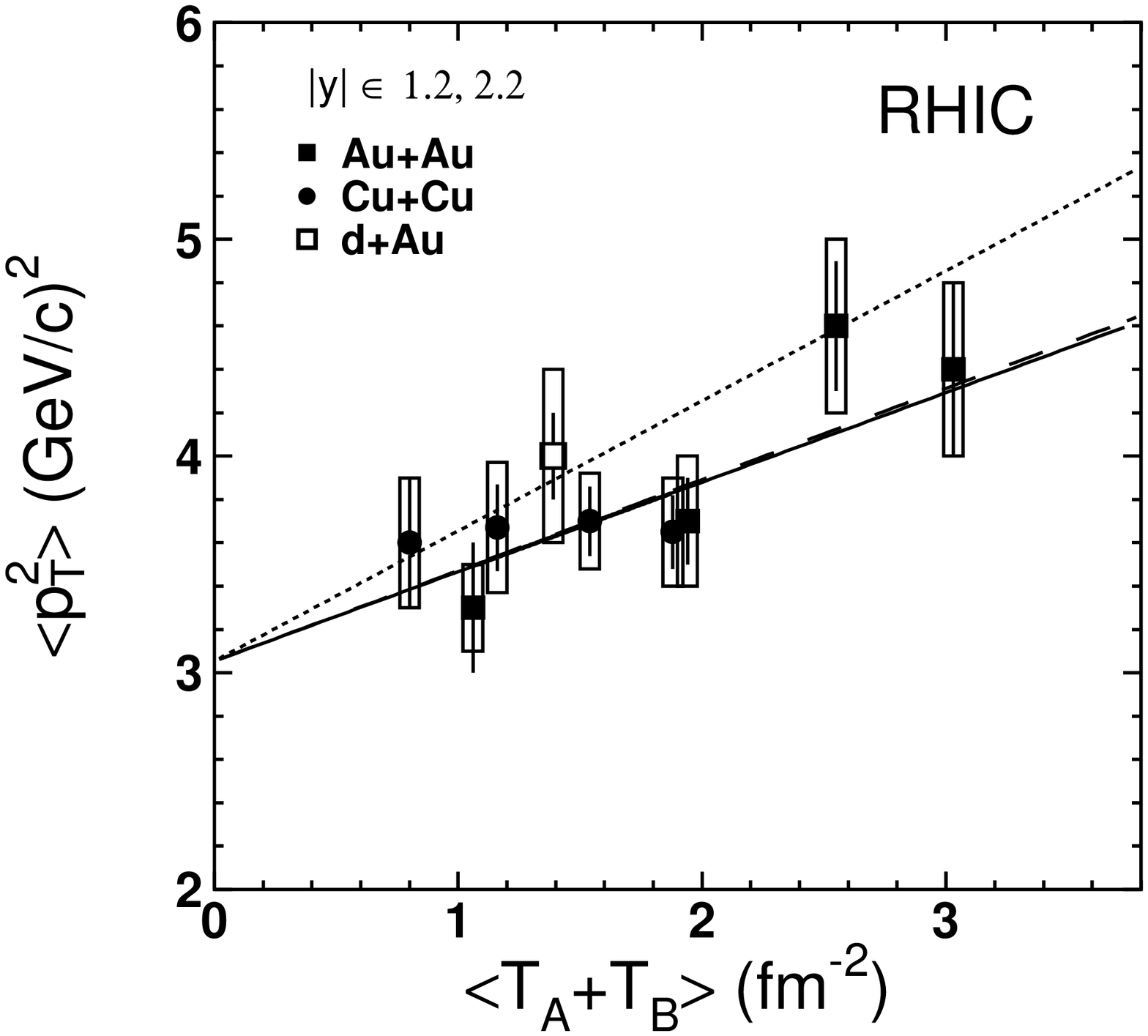}
\includegraphics[width=4.3cm]{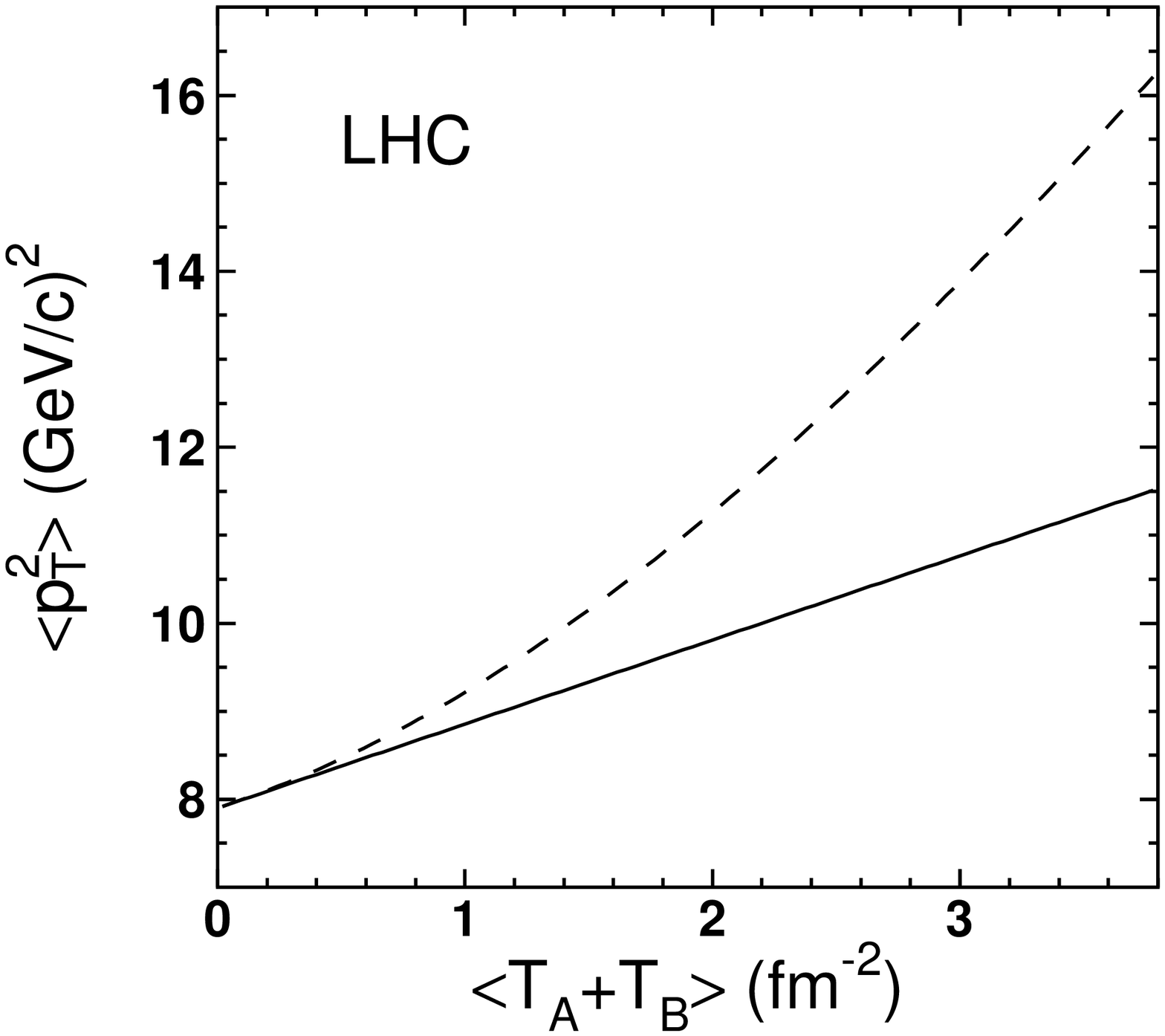}
\end{center}
\caption{\label{ta-dep} Data \cite{phenix-psi} for the mean transverse momentum squared of $J/\Psi$ produced at $\sqrt{s}=200\GeV$ an at $y=0$ (left panel)  and $y=1.2-2.2$ (middle panel) plotted as function of $\langle T_A+T_B\rangle$ defined in (\ref{520}).
Solid and dashed curves are calculated for $AA$ collision without and with the boosting effect. Same for $\sqrt{s}=5.5\TeV$ is shown in the right panel. The $d$-$Au$ point in the middle plot should be compared with the dotted curve.}
 \end{figure}

Fig.~\ref{ta-dep} presents RHIC data \cite{phenix-psi} at $\sqrt{s}=200\GeV$ for the mean $J/\Psi$ transverse momentum squared versus the mean nuclear thickness covered by the projectile gluon in the rest frame of each of the colliding nuclei, calculated at impact parameter of collisions, $\vec b$, corresponding to the measured centrality bin,
\beq
\la T_A+T_B\ra = \frac{1}{T_{AB}(b)}
\int d^2s\,\left[T_A(\vec s)+T_B(\vec b-\vec s)\right]\,T_A(\vec s)\,T_B(\vec b-\vec s),
\label{520}
\eeq
where $T_{AB}(b)=\int d^2s\,T_A(\vec s)\,T_B(\vec b-\vec s)$.
Data are compared with the prediction based on Eq.~(\ref{410}) applied to $J/\Psi$ production (see details in \cite{broad}) depicted by solid and dotted curves for $Au$-$Au$ and $d$-$Au$ collisions respectively. These curves must coincide at $y=0$ (left panel), but are different at forward rapidities (middle panel), where $x_A\neq x_B$, and we fixed rapidity at $\la y\ra=1.7$. We predict here only broadening, i.e. the slope of the curves, while the absolute value of $\la p_T^2\ra$, which is model dependent and has not been calculated so far, is treated as a fitting parameter.

We also solved Eqs.~\ref{380} applied to $J/\Psi$ production, and plotted the result by dashed curves. While at the mid rapidity (left panel) the effect boosting is sizable, although not strong, at forward rapidities (middle panel) it is hardly visible. This is a result of compensation between the effects of rising and decreasing Bjorken $x$ is the two nuclei.  

Within the rather large error bars data agree with the theoretical expectations, but cannot resolve the weak boosting effect. Extrapolated to small values of $\la T_A+T_B\ra$ all curves should meet at the value of $\la p_T^2\ra$ for $pp$ collisions. We found $\la p_T^2\ra_{pp}=3.52\pm0.3$ and $3.06\pm0.3$ at $y=0$ and $1.7$ respectively, which agree within the errors with the measured values \cite{phenix-psi}.

A much stronger boosting effect for broadening of $J/\Psi$ is expected at LHC. An example at $\sqrt{s}=5.5\TeV$ at $y=0$ is plotted in the right panel of Fig.~\ref{ta-dep}. Broadening in $AA$ collisions is enhanced up to factor three compared with $pA$ collisions at the same path length in nuclear matter.

Notice, that this kind of enhancement for broadening in $AA$ compared to $pA$ collisions was observed recently in high statistics measurement of $J/\Psi$ production in the NA60 and NA50 experiments at $E_{lab}=158\GeV$ \cite{na60-pt}.
The magnitude of broadening in nuclear collisions was found twice as big as in $pA$ measurements for the same path length in a nuclear medium.
The magnitude of the observed boosting is much larger than follows from equations (\ref{380}) at this energy,
and is probably related to another mechanism enhancing broadening in $AA$ collisions due to interaction with gluons radiated in the preceding multiple collisions \cite{prompt-gluons}. This mechanism correctly predicted the magnitude of the effect observed in the NA50/60 experiments. However, this contribution steeply falls with energy \cite{prompt-gluons}, and is negligible at the energies of RHIC and LHC.

Another observable sensitive to the saturation scale is hadron multiplicity \cite{kln}. In this case the boosting effect should lead to a jump of multiplicity in $pA$ and $AA$ collisions at the same number of participants. Indeed, such a discontinuity was observed \cite{phobos1,phobos2} in data for $Au$-$Au$ and $Cu$-$Cu$ collisions at $\sqrt{s}=200\GeV$ in comparison with multiplicity $dA$ and $pp$ collisions. The magnitude of the observed enhancement is in accord with the boosting factors presented in Fig.~\ref{results}. A detailed analysis of the data and comparison with theoretical expectations goes beyond the scope of this Letter, and will be published separately.

\section{Summary}

Nuclear targets have a larger resolution than a proton for parton distribution in the projectile hadrons. As a result, the projectile parton distribution distribution is suppressed at large $x$ and enhanced at small $x$.

The increase of projectile parton densities in the case of nuclear collisions becomes a source of enhancement of the broadening experienced by the target partons propagating through the projectile nucleus, i.e. to an increase of the saturation momentum in the beam (see Fig.~\ref{results}).
Such a mutual enhancement of the saturation scales leads to the system of reciprocal  equations (\ref{380}).

We solved the equations for central heavy ion collision and found
the saturation scale for gluon radiation to be boosted up to a factor $1.5$ at the
energy of RHIC and a factor $3$ at LHC.

The saturation scale in nuclear collisions can be probed by measuring broadening of heavy quarkonia, which is not affected by final state interaction with the created dense medium. At the energy of RHIC the boosting effect is too weak to be observed in the currently available data for $J/\Psi$ production, however it should be easily detected at the energies of LHC.

\section*{Acknowledgments}

B.K. thanks Dima Kharzeev and Raju Venugopalan for informative and
helpful discussions. This work was supported in part by Fondecyt
(Chile) grants 1090236, 1090291 and 1100287, and by DFG (Germany)
grant PI182/3-1, and by Conicyt-DFG grant No. 084-2009.

\end{document}